\documentclass[fleqn,10pt]{wlscirep} \title{Stationary bubble formation and cavity collapse in wedge-shaped hoppers} \author[]{Yui Yagisawa} \author[]{Hui Zee Then} \author[1,*]{Ko Okumura} \affil[]{Department of Physics and Soft Matter Center, Ochanomizu University, 2--1--1, Otsuka, Bunkyo-ku, Tokyo 112-8610, Japan} \affil[*]{corresponding author: okumura@phys.ocha.ac.jp} 

\begin{abstract}
The hourglass is one of the apparatuses familiar to everyone, but reveals
intriguing behaviors peculiar to granular materials, and many issues are
remained to be explored. In this study, we examined the dynamics of falling
sand in a special form of hourglass, i.e., a wedge-shaped hopper, when a
suspended granular layer is stabilized to a certain degree. As a result, we
found remarkably different dynamic regimes of bubbling and cavity. In the
bubbling regime, bubbles of nearly equal size are created in the sand at a
regular time interval. In the cavity regime, a cavity grows as sand beads fall
before a sudden collapse of the cavity. Bubbling found here is quite visible
to a level never discussed in the physics literature and the cavity regime is
a novel phase, which is neither continuous, intermittent nor completely
blocked phase. We elucidate the physical conditions necessary for the bubbling
and cavity regimes and develop simple theories for the regimes to successfully
explain the observed phenomena by considering the stability of a suspended
granular layer and clogging of granular flow at the outlet of the hopper. The
bubbling and cavity regimes could be useful for mixing a fluid with granular materials.

\end{abstract}
\begin{document}

\flushbottom\maketitle
\thispagestyle{empty}

\section*{Introduction}

Dynamics of dense granular materials is one of the challenging problems in
physics \cite{PouliquenBook,HerminghausText}. For example, recently, dynamic
aspects of jamming transitions in granular systems have emerged as an
important issue and have received considerable attention both experimentally
and theoretically
\cite{DurianPRL10,bi2011jamming,CloitreNP2011,TakeharaPRL2014,Hayakawa2015PRL}%
. Contrary to this, the flow in an hourglass or the discharge of a silo
through an orifice is a classic and familiar problem of granular dynamics, but
these phenomena still remain an issue actively discussed in recent studies. In
general, such granular flow can be continuous, intermittent
\cite{HourglassTick1993PRL,HourglassPRE1996}, or completely blocked
\cite{HopperJam2001PRL,HopperJam2006PRL}. In the continuous region, standard
views on the hourglass dynamics
\cite{HourglassNature1961,beverloo1961flow,Nedderman1992Granular,JaegerNagelBehringer96RevModPhys}
have recently been questioned via self-similar dynamics
\cite{Hourglass2012PRL}, universal scaling in a pressure transition
\cite{Hourglass2015PRL}, and independent control of the velocity and pressure
\cite{Hourglass2010PRL}. In the intermittent region, clogging transitions
under some agitation (e.g. vibration or random forces) are discussed from a
unified viewpoint \cite{CloggingSciRep2014}. In the blocked region, clogging
in tilted hoppers is discussed to clarify a phase diagram
\cite{DurianCloggingGM2010,DurianCloggingPRE2013} and jamming in a
two-dimensional hopper is explained by associating the arch at the orifice
with a random walk \cite{JammingHopperPRL2001}.

Here, we studied granular flow in an hourglass without circular symmetry (more
specifically, wedge-shaped hoppers) when a suspended granular layer could be
stabilized to a certain extent, with the aid of cohesion force between sand
grains, friction with side walls, and low permeability of the sand layer
(which leads to the pressure difference developed in the layer)
\cite{HourglassTick1993PRL,HourglassPRE1996}. As a result, we found two
spontaneous behaviors: stationary formation of bubbles and cavity collapse. We
show that these two opposite effects emerge via the interplay between
stability of the sand layer and clogging at the orifice. Note that the
clogging discussed in the present study is temporal in the sense that the flow
recovers without adding external perturbation and is governed by the cohesion
between particles unlike in the previous studies
\cite{HopperJam2001PRL,DurianCloggingGM2010,DurianCloggingPRE2013}, and thus
quite different from the clogging studied in the previous studies.

The cavity regime cannot be categorized into any previously known phases: it
is neither continuous, intermittent nor completely blocked phases. Any
systematic physical study on this original regime using a small-scale
laboratory setup have never been discussed in the physics literature. We show
that this intriguing phenomenon can be understood by the balance between the
gravitational and frictional forces.

The bubbling reported here is quite visible to a level never reported in the
literature. Although a similar phenomenon is reported
\cite{HourglassTick1993PRL,HourglassPRE1996} (a density wave or flow pattern
is also reported for much larger grains in hoppers by use of X-Ray radiography
\cite{PatternHopperPRL1989}), the bubbles are much less visible. In fact, what
are called "bubbles" in the previous study \cite{HourglassPRE1996} are quite
localized near the orifice: the "bubble" sizes are typically about 6 mm and
they "rise" at most 2-3 mm in an hourglass with the orifice of size 3.7 mm. In
contrast, in the present case, although the orifice size is smaller and is
typically 2 mm, the bubble sizes can be more than 10 mm and they can rise more
than 70 mm (they typically rise\ to the top surface of the sand layer). In
addition, in the present case, more than one bubble can coexist at the same
time, as often seen when bubbles are created not in sand but in liquid.

Since the bubbling in the present study is so clear as if it appeared in
liquid, it is possible to study the bubble size and the rising speed in
detail. As a result, we elucidate the dependence of the period of bubble
nucleation on the geometry of hourglass and clarify the importance of clogging
in the periodic motion.

Note that, although visible bubbling in a granular layer has been discussed
for fluidized beds \cite{davidson1963fluidised,davidson1985fluidization},
bubbles in the context emerge as a result of an externally imposed fluid flow.
In contrast, bubbles discussed here appear spontaneously and repeatedly in an
hourglass without any externally imposed flow.

\begin{figure}[ptb]
\begin{center}
\includegraphics[width=0.6\textwidth]{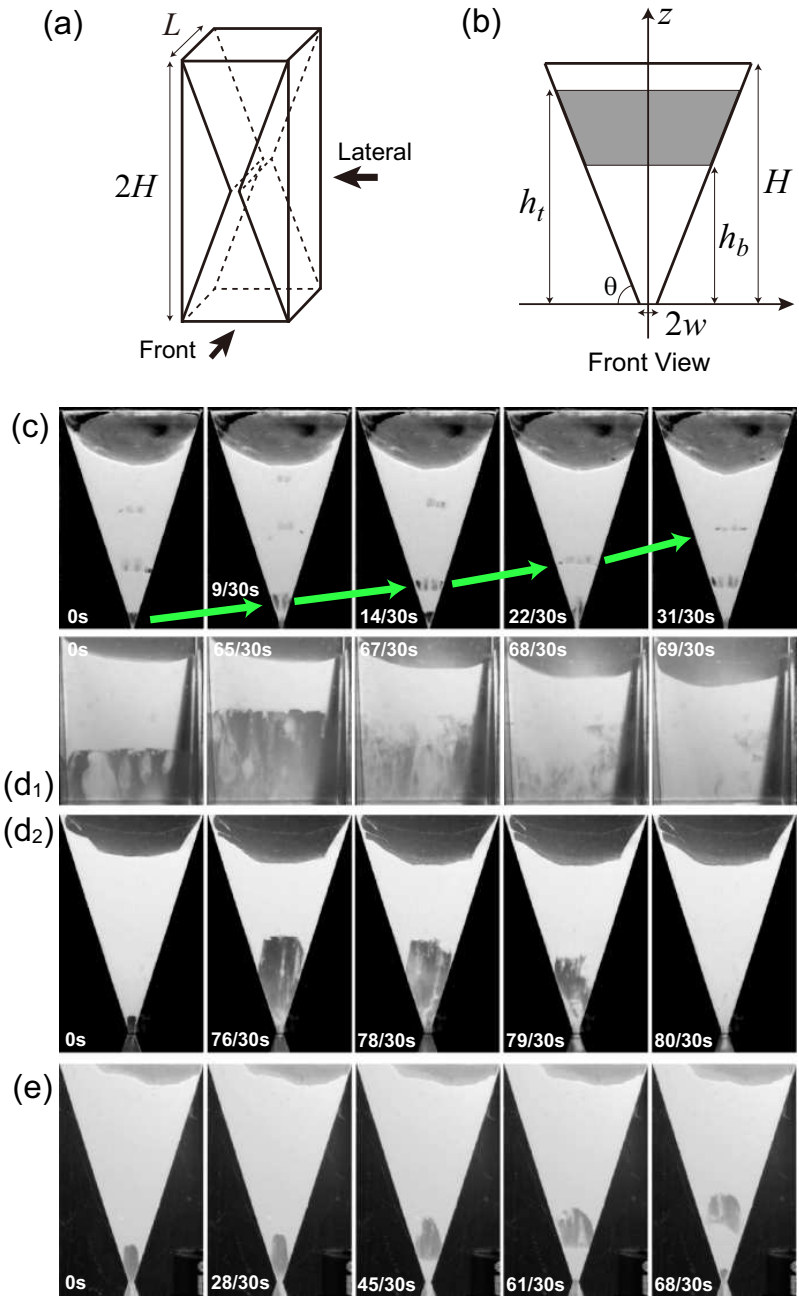}
\end{center}
\caption{(a) Illustration of an hourglass without circular symmetry. The front
and lateral sides are defined as indicated by the arrows. (b) The front view
of the upper half of the acrylic cell in (a). The position of the top and
bottom surfaces of the sand layer measured from the outlet is called $h_{t}$
and $h_{b}$, respectively. The slope is called $\alpha$ ($=\tan\theta$). (c)
Series of snapshots taken from the front side of the cell in the bubbling
regime where the diameter of glass beads $d$, the cell thickness $L$, the half
size of the outlet $w$ and the angle $\theta$ are given as $(d,L,w,\theta
)=(60,3,1,72.4)$ in the units $\mu$m, mm, mm, and deg., respectively. (d$_{1}%
$) Snapshots taken from the lateral side in the cavity regime for
$(d,L,w,\theta)=(30,56,1,72.4)$ in the same units. (d$_{2}$) Snapshots taken
from the front side in the cavity regime for $(d,L,w,\theta)=(30,56,1,72.4)$.
(e) Snapshots from the front side in the intermediate regime for
$(d,L,w,\theta)=(30,3,1,72.4)$. Corresponding movies are available for (c) and
(d$_{1}$), as Supporting Information video files.}%
\label{Fig1}%
\end{figure}

\section*{Results}

We use a transparent closed cell of height $2H$, thickness $L$, and slope
$\alpha$ (that defines the angle $\theta$ via $\alpha=\tan\theta$), as shown
in Fig.~\ref{Fig1}(a) and (b). The cell is half-filled with glass beads, which
are nearly monodisperse. The outlet width $2w$ of the cell is much larger than
the average diameter $d$ of the beads. By rotating the cell upside down, all
the glass beads in the cell that are initially located above the outlet start
falling down through the outlet due to gravity.

\subsection*{Dynamic regimes}

Depending on the parameter set, $d$, $L$, $\theta$, and $w$, the dynamics are
categorized into three regimes. (1) Bubbling regime: small air bubbles are
regularly created at the outlet and goes up to disappear, as shown in
Fig.~\ref{Fig1}(c). (2) Cavity regime: a large cavity starts to grow from the
outlet to finally collapse, as seen in Fig.~\ref{Fig1}(d$_{1}$) and (d$_{2}$).
(3) Intermediate regime: a cavity grows upwards first but later glass beads
accumulate near the clogged outlet, leading to a "rising cavity," as in
Fig.~\ref{Fig1}(e).

\begin{figure}[ptb]
\begin{center}
\includegraphics[width=0.8\textwidth]{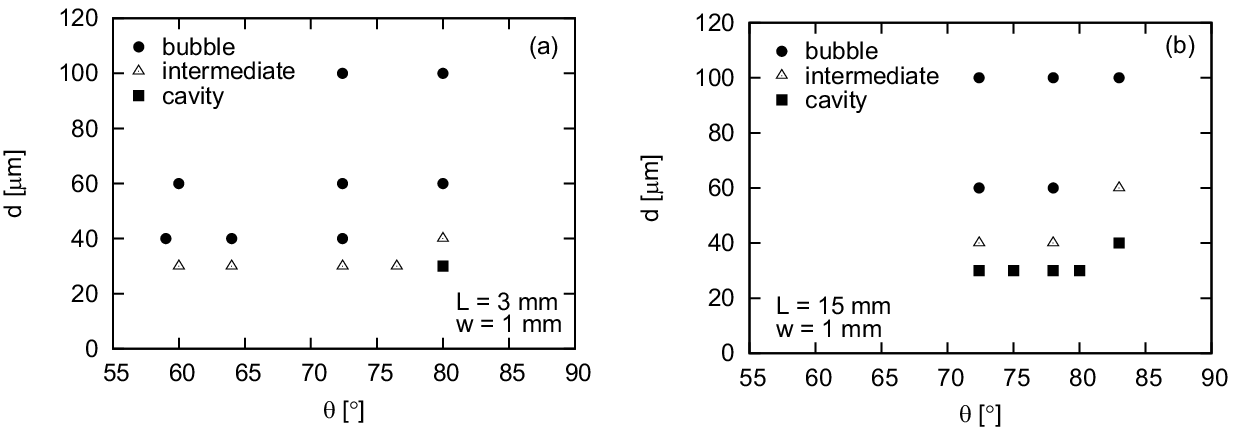}
\end{center}
\caption{Phase diagram as a function of $\theta$ and $d$ obtained from thin
(a) and thick (b) cells. }%
\label{Fig2}%
\end{figure}

The phase diagrams in Fig.~\ref{Fig2} can be physically understood by
considering the following points:

(1) The dynamic regime is determined as a result of competition of the
stability of the bottom free surface of the sand layer and the clogging at the
orifice. A cavity can keep growing only when the free surface is stable and
only when clogging is difficult to occur. Bubbling occurs under the opposite
conditions (low stability and easy clogging): a bubble looks as if to "rise
up" in the granular medium, as a result of the collapse of the top part (due
to low stability) followed by the accumulation of sand (due to clogging).

(2) The stability is increased when particles are fine. This is because the
cohesive force between particles becomes large when the sand is fine. This is
well-known at least empirically and explained frequently in terms of capillary
bridges between sand particles originating from a tiny amount of water located
at the contacts between grains \cite{PouliquenBook,HerminghausText}. The ratio
of the capillary force $k\gamma d$ to the gravitational force $\rho gd^{3}$
for a particle of diameter $d$, which scales as $k(l/d)^{2}$, becomes larger
when particle becomes small. Here, $\gamma,\rho$ and $g$ are the surface
tension of water, the density of the particle, and the gravitational
acceleration, defining the capillary length $l=\sqrt{\gamma/(\rho g)}$. In the
atmospheric humidity, the numerical coefficient $k$, which reflects the amount
of water at the contacts of particles, would be very small. However, for
example, when $d$ is 30 $\mu$m, the ratio $k(l/d)^{2}$ could become comparable
to unity because $l$ is about 3 mm: for small particles, the cohesive force
can excel the gravitational force. Another reason for the high stability of
fine particles is the small permeability of the sand layer formed by fine
grains. This will help create pressure difference in the sand layer,
contributing the stability. This point will be discussed in more detail in the
next two sections.

(3) The clogging occurs easily when the slope is small and when the cell is
thin. Easy clogging for small slopes is reasonable: the smaller the slope is,
the more concentrated towards the outlet the falling sand beads are. Easy
clogging for thin cells (small $L$) is understandable from snapshots in
Fig.~\ref{Fig1} (d$_{1}$), as explained as follows. In each snapshot, we can
observe streams of falling sand whose width and numbers are changing with time
(for example, in the left-most snapshot in Fig.~\ref{Fig1} (d$_{1}$) we
recognize 4 streams three of which have almost the same width while the
left-most stream is the widest). When the cell thickness is small and
comparable to a typical width of such streams, the flow flux averaged by the
orifice area is essentially the same with the flow flux of the stream.
However, when the cell thickness is large and several streams are observed
with some spacing between them as in Fig.~\ref{Fig1} (d$_{1}$), the average
flux at the orifice is smaller than the flow flux of the streams. As a result,
we expect that clogging is more difficult to occur in a cell of large
thickness because of the smaller average flux (Note that the stability is not
completely independent from the clogging: The flux is correlated with the
stability of the free surface so that high (low) stability implies that
clogging is difficult (easy)). In summary, clogging is easy to occur for thin
cells whose slope is small.

On the basis of the above points (1)-(3), the phase diagrams in
Fig.~\ref{Fig2} can be physically understood: The cavity regime should be
observed only when the diameter $d$ of sand is small (for strong cohesion and
low permeability), the cell thickness $L$ is large, and the angle $\theta$ is
large (for clogging to be difficult), whereas the bubbling regime tends to be
seen under the opposite conditions, i.e., when $d$ is large, $L$ is small, and
$\theta$ is small. These tendencies are clearly confirmed in Fig.~\ref{Fig2}
for $w=1$ mm.

\subsection*{Bubbling regime}

In the bubbling regime, the $n$th bubble is created as follows. First, a
cavity is nucleated at the outlet at time $t=t_{\text{cav}}^{(n)}$ and, then
the cavity grows as beads keep falling ($t_{\text{cav}}^{(n)}$ is the
nucleation time of the $n$th cavity). At time $t=t_{\text{bub}}^{(n)}$,
clogging occurs at the outlet, closing the outlet: this is the moment of
creation of a bubble ($t_{\text{bub}}^{(n)}$ is the creation time of the $n$th
bubble). The $n$th bubble thus created rises up in the granular medium
upwards, whereas after a short waiting time $t_{\text{w}}^{(n)}$ (i.e., at
time $t_{\text{cav}}^{(n+1)}=t_{\text{bub}}^{(n)}+t_{\text{w}}^{(n)}$) the
next $(n+1)$th cavity is nucleated at the clogged part at the bottom, which
leads to the next bubble.

\begin{figure}[ptb]
\begin{center}
\includegraphics[width=0.8\textwidth]{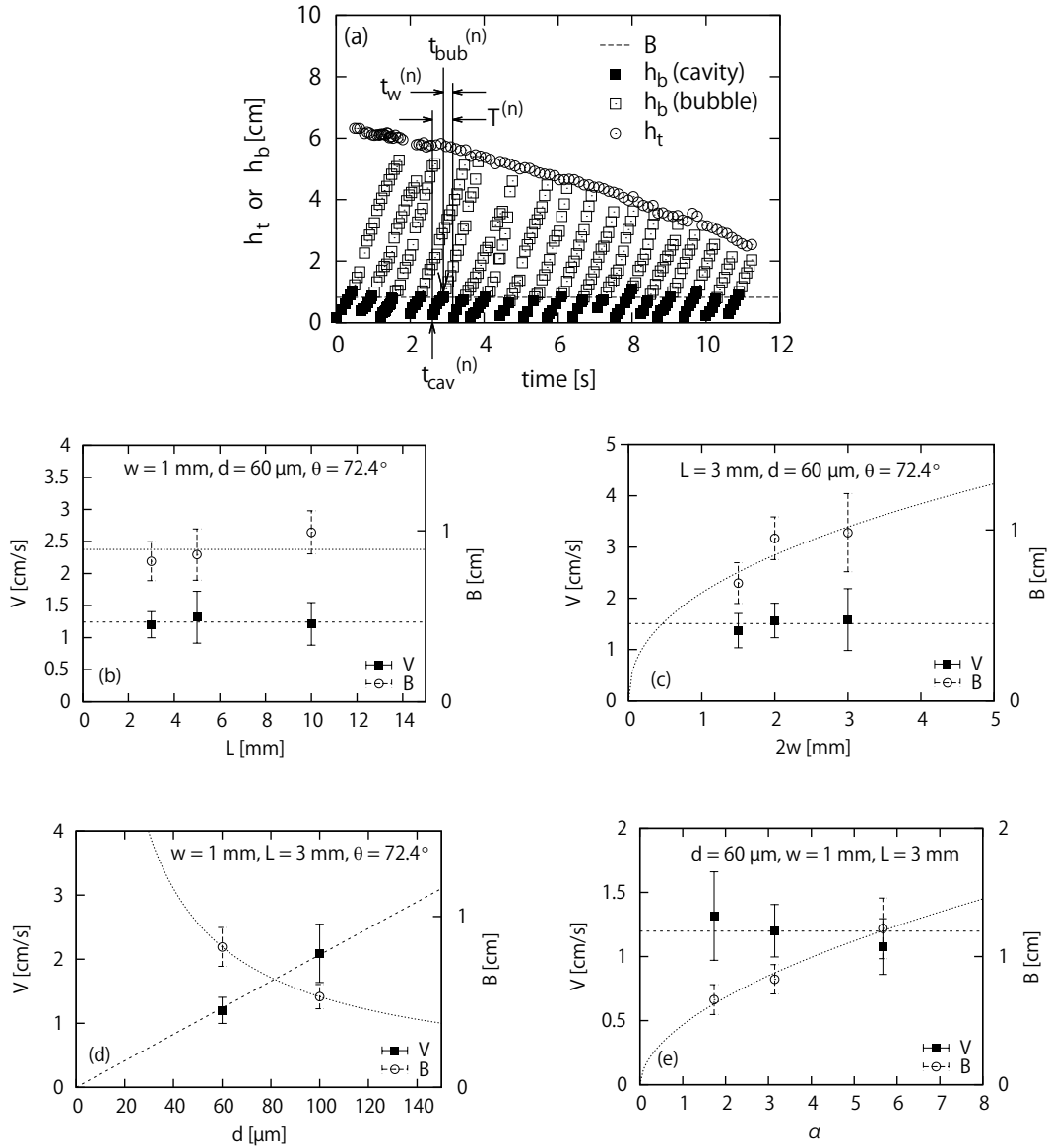}
\end{center}
\caption{(a) $h_{t}$ and $h_{b}$ (see Fig.~\ref{Fig1}(b)) vs time in the
bubble regime for $d=60$ $\mu$m, $L=3$ mm, $w=1$ mm, and $\theta=72.4^{\circ}$
The quantities $t_{\text{cav}}^{(n)}$, $t_{\text{bub}}^{(n)}$, and
$t_{\text{w}}^{(n)}$ are the nucleation time of the $n$th cavity, the creation
time of the $n$th bubble, and the waiting time for the ($n+1)$th cavity,
respectively. The horizontal dashed line shows the average value $B$ of the
bubble size $B^{(n)}$ at the time of bubble creation. (b) $V$ and $B$ vs $L$
for a given $w,d$, and $\theta$. (c) $V$ and $B$ vs $2w$ for a given $L,d$,
and $\theta$. (d) $V$ and $B$ vs $d$ for a given $w,L$, and $\theta$. (d) $V$
and $B$ vs $\alpha$ $(=\tan\theta$) for a given $L,d$, and $w$. The fitting
curves for (c), (d), and (e) are given as, $B=(0.63\pm0.15)\times
(2w)^{0.43\pm0.28}$, $B=27d^{-0.85}$ and $V=(0.021\pm0.000)\times d$, and
$B=(0.47\pm0.06)\times\alpha^{0.54\pm0.09}$, respectively.}%
\label{Fig3}%
\end{figure}

To quantify this dynamics, the positions of the top and bottom surfaces of the
sand layer, $h_{t}$ and $h_{b}$ (defined in Fig.~\ref{Fig1}(b)), are plotted
in Fig.~\ref{Fig3}(a) for a parameter set in the bubbling regime. The position
of the bottom is marked by squares: Filled and open squares correspond to the
data before and after the clogging at $t=t_{\text{bub}}^{(n)}$, respectively.
In other words, filled and open squares stand for the top surface of a cavity
and that of a bubble, respectively.

Figure~\ref{Fig3}(a) shows a repeated or regular dynamics: the time interval
between consecutive bubble creations, the rising speed of bubbles, and the
initial bubble size are almost constant. As seen in Fig.~\ref{Fig3}(a), the
cavity nucleation (at time $t=t_{\text{cav}}^{(n)}$) and the bubble formation
(at time $t=t_{\text{bub}}^{(n)}$) occur regularly and the interval defined as
$T^{(n)}=t_{\text{cav}}^{(n+1)}-t_{\text{cav}}^{(n)}$ is almost a constant
$T$. The growing speed of the $n$th cavity $V^{(n)}=(dh/dt)^{(n)}$ defined for
the interval from $t_{\text{cav}}^{(n)}$ to $t_{\text{bub}}^{(n)}$ is also
almost a constant $V$, whereas the velocity $V^{(n)}$ is comparable to the
rising speed of the $n$th bubble defined after $t_{\text{bub}}^{(n)}$. This
means that the initial bubble size $B^{(n)}$, defined as $B^{(n)}%
=h_{b}(t_{\text{bub}}^{(n)})$, is also a constant $B$, as seen in
Fig.~\ref{Fig3}(a), whereas the following equation holds:
\begin{equation}
B^{(n)}=V^{(n)}(T^{(n)}-t_{\text{w}}^{(n)})=V^{(n)}(t_{\text{bub}}%
^{(n)}-t_{\text{cav}}^{(n)}). \label{eq0}%
\end{equation}

The $n$ - independence of the growing speed $V^{(n)}=(dh_{b}/dt)^{(n)}$ is
interpreted as follows. This quantity essentially corresponds to the
collapsing speed of the bottom surface (at height $z=h_{b}$) of the suspended
granular layer, which can be determined as follows. By introducing a life time
$\tau$ of a sand particle after the moment the sand particle is exposed to the
sand-air interface (positioned at height $z=h_{b}$), we simply obtain
\textit{the growing speed of a cavity or the rising speed of a bubble},
\begin{equation}
V=dh_{b}/dt=d/\tau. \label{eq1}%
\end{equation}
As seen in Fig.~\ref{Fig3}(a), the slope $(dh_{b}/dt)^{(n)}$ is almost
independent of $n$ and $h_{b}$ (although the slope has a slight tendency to
increase with $h_{b}$). This means that the life time $\tau$ and the rising
speed of bubbles are also independent of $n$ and $h_{b}$ (although $\tau$ has
a slight tendency to decrease with $h_{b}$ and the average rising speed of
bubbles has a slight tendency to increase with $h_{b}$). The physical nature
of $\tau$ will be explored in Discussion.

The $n$ - independence of the initial bubble size $B^{(n)}$ is interpreted as
follows. We introduce the two dimensional volume fraction of sand beads in the
layer $\phi$, which is practically assumed to be the random closed packing
value $\phi_{c}$ as an approximation. The volume flux created at the bottom
surface of the sand layer, $Q_{b}$, is given as $Q_{b}=\phi SV$, where $S$ is
the surface area of the free surface at $z=h_{b}$ (note that in the bubbling
regime, $L$ is smaller than the typical width of streams of flowing sand
mentioned above) and thus given by $S=Lh_{b}/\alpha$ (when $w$ is small). The
maximum flux at the outlet $Q_{0}$ is estimated as $Q_{0}=\phi_{c}S_{0}V_{0}$,
with the section area $S_{0}=2wL$ of the cell at the outlet ($z=0$) (see
Fig.~\ref{Fig1}(b)) and the velocity at the outlet $V_{0}$. This velocity is
determined by $V_{0}^{2}=2g(h_{b}-\Delta)$, where $g\Delta$ expresses a small
energy dissipation (per unit mass) for sand beads falling off from the surface
after gliding on the cell wall of the slope $\alpha$. By noting that
$Q_{b}\simeq h_{b}$ and $Q_{0}\simeq\sqrt{h_{b}}$ ($\Delta$ is generally small
in the experiments), we see that, just after the nucleation of a cavity, the
clogging does not occur, because $Q_{b}<Q_{0}$ (i.e., $h_{b}<\sqrt{h_{b}}$)
for small $h_{b}$. Then, the critical condition for clogging is given by
$Q_{b}=Q_{o}$. The value of $h_{b}$ obtained by solving this critical
condition in favor of $h_{b}$ corresponds to the bubble size $B$ (This is
because the size of $h_{b}$ at the creation of the bubble is the definition of
the initial size of the bubble, $B$, as defined above). From this clogging
condition and Eq.~(\ref{eq1}), we obtain%
\begin{equation}
B=u\beta\text{ \ with \ }\beta=g(2\alpha w\tau/d)^{2}\label{eq2}%
\end{equation}
where $u=1+\sqrt{1-2\Delta/\beta}$ with $\Delta\ll\beta$. This explains why
the horizontal dashed line in Fig.~\ref{Fig3}(a) gives a well-defined average
value $B$, that is, why $B^{(n)}$ is almost independent of $n$. Since the
waiting time $t_{\text{w}}^{(n)}$, which corresponds to the reorganization
time for an once packed granular layer formed near the outlet to destabilize
(see below for more details), is expected to be $n$-independent, the
$n$-independence of $V^{(n)}$ and $B^{(n)}$ justify the $n$-independence of
$T^{(n)}$ [see Eq.~(\ref{eq0})].

The agreement of the above theory with experiment is seen in different ways in
Figure~\ref{Fig3}(b)-(e), which supports Eqs.~(\ref{eq1}) and (\ref{eq2}).
Note that $B$ is predicted as an increasing function of $\alpha w/d$ from
Eq.~(\ref{eq2}) as understood from the approximate expression $B\simeq2\beta$
obtained in the limit $\Delta\ll\beta$. Then all the following behaviors shown
in Figure~\ref{Fig3}(b)-(e) are consistent with Eqs.~(\ref{eq1}) and
(\ref{eq2}), if the dependence of $\tau$ on $d$ is relatively weak. (1)
Figure~\ref{Fig3}(b) shows that $V$ and $B$ are independent of $L$. (2)
Figure~\ref{Fig3}(c) shows that $V$ is independent of $w$ and that $B$
increases with $w$. (3) Figure~\ref{Fig3}(d) shows that $V$ increases with $d$
and that $B$ decreases with $d$. (4) Figure~\ref{Fig3}(e) shows that $V$ is
independent of $\alpha$ and that $B$ increases with $\alpha$.

\begin{figure}[ptb]
\begin{center}
\includegraphics[width=0.6\textwidth]{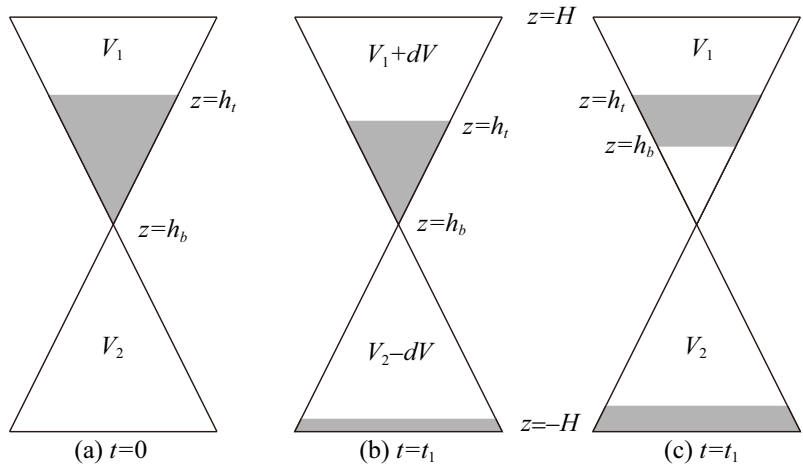}
\end{center}
\caption{Creation and non-creation of air pressure difference in the chambers
of an hourglass. The shaded areas stand for the regions occupied by the sand,
which is here considered as completely impermeable. In the change from (a) at
the time $t=0$ to (b) at the time $t=t_{1}$, the pressure difference is
created because of the changes in volume of the air in the chambers (from the
volumes $V_{1}$ and $V_{2}$ to $V_{1}+dV$ and $V_{2}-dV$). In the changes from
the state in (a) to the state in (c), the pressure difference is not created,
corresponding (nearly) to the case in the cavity regime.}%
\label{Fig4A}%
\end{figure}

The importance of a weak and localized pressure difference created near the
orifice was clearly shown in the intermittent regime
\cite{HourglassTick1993PRL,HourglassPRE1996}. They showed the existence of the
active phase in which the flow through the orifice is maintained and the
pressure difference increases with the increase in $h_{b}$ till the bubble
disappears. This active phase is followed by the inactive phase in which the
flow through the orifice is stopped and the pressure difference decreases till
the creation of the next cavity. During the active phase, because of the
transfer of sand from the upper to lower chamber, the pressure in the sand
layer near the orifice becomes larger than the pressure in the upper chamber
due to the compression and expansion of air in the lower and upper chambers,
respectively, provided that the chambers are closed (Note, however, this is
true only when $h_{t}$ decreases as $h_{b}$ increases; see Fig.~\ref{Fig4A}).
This pressure difference helps stabilize the bottom of the sand layer, leading
to a complete stop of the flow, i.e., to an initiation of the inactive phase.
During the inactive phase the pressure differences thus created is gradually
reduced by the permeation of air in the sand layer till the bottom of the sand
destabilizes again, leading to an intermittent dynamics. They developed a
theory describing the duration of inactive phase $T_{i}$ on the basis of the
Darcy law on the permeation of viscous fluid in porous materials and confirmed
the theory by experiments. However, they did not give a theory describing the
duration of the active phase $T_{a}$ although they indicated the importance of
clogging, by explicitly saying \cite{HourglassPRE1996} "the plug was created
when the flux of particles coming from the free-fall arch (corresponding to
the bottom surface of the sand layer at $z=h_{b}$) was too large to pass
through the orifice rapidly." Note here the following points for this series
of study \cite{HourglassTick1993PRL,HourglassPRE1996}: (1) They claimed that
the period $T_{i}+T_{a}$ was quite robust and independent of the grain size
and explained the characteristic order of the period by the Darcy dynamics in
the first series of study \cite{HourglassTick1993PRL}; (2) However, in the
second series of study \cite{HourglassPRE1996} the view on the robustness was
modified and they suggested the importance of clogging in the active phase as
above but without developing a theory.

Bubbling in the present study and in the previous studies
\cite{HourglassTick1993PRL,HourglassPRE1996} are quite different. The size of
bubbles is much larger and the life time of bubbles is much longer in the
present study. A marked difference is that bubbles in the present cases never
disappears before they reach the top of the sand layer $h_{t}$. In addition,
more than one bubble can coexist in the sand layer in the present case.

However, we consider that the pressure difference is still important in the
present bubbling regime, which is an intermittent regime as in the previous
studies \cite{HourglassTick1993PRL,HourglassPRE1996}, although detailed
effects should be rather different (see below). For example, the period
$T_{i}$ in the previous study, in which period the pressure difference is
reduced, corresponds to the waiting time $t_{\text{w}}$ in the present study.
Our theoretical arguments based on the clogging condition given above provides
an estimate for the counterpart of $T_{a}$ in which period the pressure
difference is increased, whereas any explicit estimates are not available in
the previous studies \cite{HourglassTick1993PRL,HourglassPRE1996}. The
counterpart of $T_{a}$ is here defined as $t_{a}\equiv t_{\text{bub}}%
^{(n)}-t_{\text{cav}}^{(n)}$, through the relation $t_{a}\simeq B/V$. This
expression implies that the time $t_{a}$ increases with $\alpha$ and $w$ and
decreases with $d$ (see Eqs.~(\ref{eq1}) and (\ref{eq2})). We have confirmed
that this is indeed the case in our experiments on the basis of the data shown
in Fig.~\ref{Fig3}(b)-(e).

Note that the dependence of $t_{a}$ on $d$ in the present study is the
opposite to that of $T_{a}$ on $d$: $t_{a}$ decreases with $d$ whereas $T_{a}$
increases with $d$. This is a clear experimental fact while the dependence of
$t_{a}$ on $d$ is explained by the present theory. This suggests that although
the physical mechanisms of bubbling in the present and previous studies are
conceptually similar but the details are quite different.

\subsection*{Cavity regime}

\begin{figure}[ptb]
\begin{center}
\includegraphics[width=0.8\textwidth]{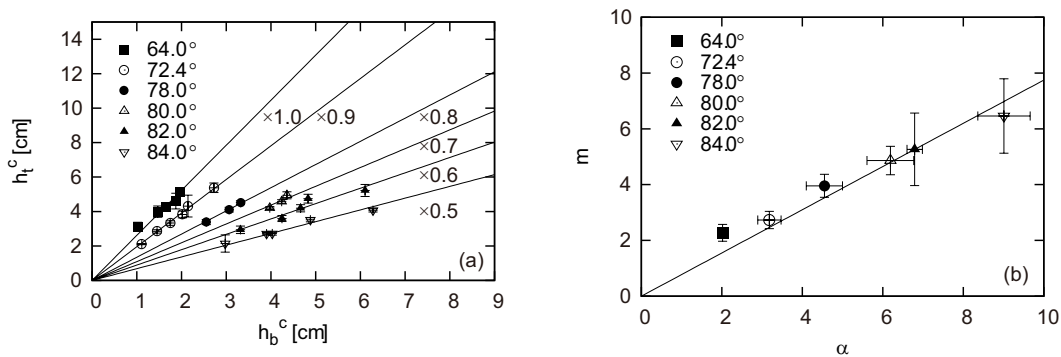}
\end{center}
\caption{(a) $h_{t}^{c}$ vs $h_{b}^{c}$, i.e., $h_{t}$ vs $h_{b}$ (See
Fig.~\ref{Fig1}(b)) at the moment of the collapse of a cavity in the cavity
regime, for various $\theta$, i.e., $\alpha=\tan\theta$ with $L=56$ mm, $2w=2$
mm and $d=30$ $\mu$m. The data for a fixed $\theta$ (or $\alpha$) is on a
straight line as predicted. To avoid overlap, the values of $h_{t}^{c}$ in
each series are scaled as suggested on the plot. (b) $m$ vs $\alpha$. The
quantity $m$ is here obtained experimentally from the slope of the line
fitting the $h_{t}^{c}-h_{b}^{c}$ plot (without the rescaling for avoiding the
overlap). The data are on the straight line as predicted, with the slope
predicts the friction coefficient $\mu$ between sand particles and cell
walls.}%
\label{Fig4}%
\end{figure}

In Fig.~\ref{Fig4}(a), the sudden collapse of the cavity is quantified by
plotting $h_{t}$ as a function of $h_{b}$ at the moment of collapse; the
critical values are respectively denoted $h_{t}^{c}$ and $h_{b}^{c}$ in the plot.

This linear relation $h_{t}^{c}\simeq h_{b}^{c}$ for various conditions shown
in Fig.~\ref{Fig4}(a) can be explained by the condition that the cavity
collapse occurs when gravitational force $F_{g}$ acting on the sand layer
exceeds friction force $F_{\text{fric}}$ acting on the layer from the cell
walls. The force $F_{g}$ is simply estimated by $\rho gL[\left(  h_{t}%
^{c}\right)  ^{2}-\left(  h_{b}^{c}\right)  ^{2}]/\alpha$ with $\rho$
representing the density of the granular layer. The force $F_{\text{fric}}$ is
estimated as $\mu\rho gL(h_{t}^{c}-h_{b}^{c})^{2}$ when $L\gg H$ as shown in
Methods. When a cavity starts growing, the condition $F_{\text{fric}}>F_{g}$
holds. As the cavity becomes larger, both forces decrease but the
gravitational force decreases more slowly. As a result, the condition
$F_{\text{fric}}=F_{g}$ is satisfied in the end, resulting in the collapse of
the cavity. This last condition for $L\gg H$ gives%
\begin{equation}
h_{t}^{c}=\frac{m+1}{m-1}h_{b}^{c} \label{eq3}%
\end{equation}
with
\begin{equation}
m=\mu\tan\theta=\mu\alpha\label{eq4}%
\end{equation}
This equation explains the linear relation shown in Fig.~\ref{Fig4}(a) (All
the data in Fig.~\ref{Fig4} are obtained for $L=56$ mm, which satisfies the
necessary condition $L\gg H$ for Eq.~(\ref{eq3}) to be valid). From the slope
of the linear fitting line that should be equal to $(m+1)/(m-1$), we obtain
$m$ as a function of $\alpha$, as shown in Fig.~\ref{Fig4}(b). As predicted in
Eq. (\ref{eq4}), $m$ is linearly dependent on $\alpha$, from which the
friction coefficient is obtained as $\mu=0.775\pm0.032$. This value also
justifies our theory because this value is a reasonable value as a friction coefficient.

The effect of pressure gradient discussed for an intermittent regime in the
previous studies \cite{HourglassTick1993PRL,HourglassPRE1996} may play an
essential role also in the cavity regime for the stabilization of the
suspended sand layer although the cavity regime is not an intermittent regime.
However, the collapse condition given in Eq.~(\ref{eq3}) is still
predominantly determined not by the pressure difference but by the friction in
the present cavity regime. There are several reasons for this as follows:

(1) The pressure difference that would be accumulated if the sand layer were
completely impermeable seems to be rather small. This is because, during the
cavity formation in which the height $h_{b}$ is increasing, the height $h_{t}$
is almost fixed to the initial position, as illustrated in Fig.~\ref{Fig4A}%
(c). Note that if $h_{t}$ is completely fixed and the sand layer is completely
impermeable, the pressure difference should not be created at all. This is
because, in such a case, as illustrated in Fig.~\ref{Fig4A}, there is no
change in volume for the air in the space above the sand layer (i.e., in the
region, $H>z>h_{t}$), and no volume change occurs also for the air in the
space below the sand layer and above the sand accumulated at the bottom of the
lower chamber (i. e., in the region, $h_{b}>z\gtrsim-H$): air is neither
compressed nor expanded.

(2) Even if a small pressure difference is created as suggested in (1), the
pressure difference seems to be considerably reduced in reality during the
formation of the cavity. This is experimentally supported in two ways. (a)
Careful observation near the top surface of the sand layer (at the position
$z\gtrsim h_{t}$) reveals that sand particles are vigorously ejected upwards
because of air flow coming out from the sand layer, which obviously relaxes
the pressure gradient. (b) As soon as $h_{b}$ approaches the critical value,
the cavity collapses without any observable delay; If the sand layer were
predominantly supported by the pressure differences accumulated before
reaching the critical state, the collapse would occur only after a delay time
in which the pressure gradient would be reduced to some extent by the
permeation of air through the sand layer; In fact, in the previous work
\cite{HourglassPRE1996} in which the effect of pressure gradient is important,
they observed such a waiting time $T_{i}$ of the order of a few seconds. The
effect of air permeation during the cavity formation, described in (a),
obviously helps stabilize the sand layer because of the viscous drag due to
air, especially because the permeability becomes smaller when the grain size
is small. However, the fact that the pressure gradient seems almost completely
relaxed by the time of the collapse (if otherwise, the delay time should be
noticeable because of the small permeability), as discussed in (b), implies
that the collapse condition cannot be strongly affected by the pressure gradient.

(3) The collapse condition governed by the gravitational and friction force
can be dimensionally justified. The two forces acting on the suspended
granular layer are given both as $\rho gLH^{2}$ as implied above because
$h_{t}$ and $h_{b}$ are roughly of the same order of magnitude as $H$. These
estimates elucidate why these two forces can balance with each other to give
the collapse condition. On the contrary, the force on the layer that
originates from the pressure difference between two chambers scales as
$cP(\Delta V/V)A$, where $\Delta V$ is very small as implied in (1) and the
numerical coefficient $c$ is also very small as suggested in (2). Here, $P$ is
the initial pressure of the chambers and $A$ is a characteristic area of the
layer, whereas $V$ and $\Delta V$ are a characteristic volume of chambers and
changes in them, respectively. Since $\Delta V/V$ and $A$ roughly scale as
$\delta/H$ and $LH$, respectively, the ratio of the pressure force to the
gravitational or frictional force scales as $c(P/\rho gH)(\delta/H)$. Here,
$\delta$ is the downwards shift of $h_{t}$ during the cavity formation, which
produces the pressure gradient. As mentioned in (1), $\delta$ is significantly
small in the cavity regime. Thus, the ratio $(P/\rho gH)(\delta/H)$ is
typically of the order of unity. This justifies why the pressure effect is
less dominant because $c$ is very small due to the effect of relaxation
described in (2).

(4) The two independent checks for the agreement between theory and experiment
shown in Fig.~\ref{Fig4}(a) and (b) strongly support that the collapse of the
cavity is properly described by the balance between the gravitational and
frictional forces. If the collapse condition is governed not by the friction
force but by the pressure difference, we expect that neither Eq. (\ref{eq3})
nor Eq. (\ref{eq4}) are correct. In reality, both Eq. (\ref{eq3}) and Eq.
(\ref{eq4}) are well confirmed as shown in Fig.~\ref{Fig4}(a) and (b), respectively.

\section*{Discussion}

At this point, we discuss crude estimates for $\tau$, although a better
physical understanding of the life time $\tau$ possibly related to a
reorganization time of granular structure will be an important issue to be
resolved in the future. One possibility is based on the free fall motion of a
grain: this gives $\tau\sim\sqrt{d/g}$. Another possibility is based on
viscous drag by the capillary bridges at the contacts: writing the balance
between the viscous and gravitational force as $\eta U/d^{2}\sim\rho g$, with
the viscosity of water being $\eta$ and with the characteristic velocity $U$
scaling as $d/\tau$, we obtain $\tau\sim\eta/(\rho gd)$. Both crude estimates
give similar orders of magnitude: $\tau\sim$ ms. This is comparable to $\tau$
estimated by $V$ and $d$ on the basis of Eq.~(\ref{eq1}). In addition, both
estimates with Eq.~(\ref{eq1}) and Eq.~(\ref{eq2}) are consistent with
qualitative behaviors of $V$ and $B$ in Fig.~\ref{Fig3}(d).

While humidity affects $\tau$ as indicated above, the precise control of
humidity is experimentally difficult. For example, the data in Fig.~\ref{Fig3}%
(a) can be obtained within approximately 10 seconds, so that we observe a well
defined $V,B$, and $T$ in Fig.~\ref{Fig3}(a). However, the data in
Fig.~\ref{Fig3}(b)-(d) cannot be obtained in such a short duration and in fact
obtained within several hours in the same room. Because of this non-precise
control of humidity, we should be satisfied not by a quantitative but by
qualitative agreement between theory and experiment in Fig.~\ref{Fig3}(b)-(d).

However, we expect the humidity dependence is not too strong. This is because
of the following reasons: (1) The bubbling regime appears for larger grains so
that the effect of cohesion should be relatively weak in the bubbling regime;
(2) The qualitative agreement between theory and experiment in the bubbling
regime was obtained, in spite of the non-precise control of humidity; (3) If
$d$-dependence of $\tau$ were strong, it could change qualitative behaviors of
$V$ and $B$ in Fig.~\ref{Fig3}(d).

The reason we observed bubbling and cavity collapse in a hopper that are quite
visible to a level never reported in the literature may be the following. The
key factor that we successfully observed spontaneous bubbling and cavity
collapse in a hopper is the stabilization of a suspended granular layer. When
the bubbling regime is observed, the cell thickness is relatively small and
the cell is quasi-two dimensional. This helps stabilize a suspended layer
because the layer is sandwiched in a small gap between the cell walls, to
create a cavity at the outlet, which is soon closed by clogging. When the
cavity collapse is observed, the cell thickness is relatively large. In this
case, the cohesion force between small sand particles helps stabilize a
suspended layer, to create a cavity at the outlet, which is not soon closed
because clogging is harder to occur when the cell thickness is large. These
stabilization conditions have not been well satisfied in most of the previous
physical studies.

However, a phenomenon similar to the one in our bubbling regime are reported
\cite{HourglassTick1993PRL,HourglassPRE1996} as already mentioned, in which
the importance of the pressure gradient is highlighted. In this respect, the
present study provides the following perspective. The pressure gradient is
especially important in intermittent regimes, which includes the bubbling
regime of the present study. In intermittent regimes, we can recognize two
phases: the active phase in which the flow through the orifice is maintained
and the inactive phase in which the flow is stopped. At least conceptually
(although the details are dependent on experimental parameters such as the
geometry of hoppers and the size of grains), the characteristic time scale for
the inactive phase is described by Darcy law for the permeation of viscous
fluid in a porous medium as discussed previously
\cite{HourglassTick1993PRL,HourglassPRE1996}, whereas the characteristic time
for the active phase is characterized by the condition of clogging as
discussed in the present paper. Note that in the cavity regime of the present
study, which is not an intermittent phase, the pressure gradient may be
important for the stabilization of the suspended sand layer but it does not
govern the condition of the collapse of the cavity.

The spontaneous bubbling and cavity collapse we discussed in this study may be
useful for mixing a fluid and powder in industrial applications. This is
expected because forced fluidization processes are essential for applications,
such as petroleum refining and biomass gasification, in which bubbles created
in fluidized bed reactors affect the efficiency
\cite{BubbleEffectsFluidizedBed1981,FluidizeExpRev1994}. A merit of the
bubbling and cavity collapse discussed in the present study is that these
phenomena occur spontaneously due to gravity and there is no need to pump a
fluid inside a granular medium: efficient or violent mixing (by virtue of the
bubbling or cavity collapse, respectively) is repeatedly achieved just by
rotating a container. In other words, the fluidized bed reactor is driven by
an externally imposed air flow, whereas the present bubbling and cavity
regimes are driven spontaneously by gravity.

To conclude, the present results reveal novel aspects of granular dynamics
when a suspended granular layer becomes stabilized to a certain extent and the
interplay of the stabilized granular bed and clogging of the granular flow is
important, which could open new avenues of research within granular physics.
We have shown (1) that bubbling and cavity emerge as a result of the interplay
of the clogging and the stability, (2) that the rising speed of a bubble
(comparable to the growing speed of a cavity), the initial bubble size and the
constant formation of bubbles are understood by introducing a life time for a
sand particle exposed to the interface of a suspended granular layer and a
clogging condition resulting from the competition between two characteristic
flow rates, and (3) that the sudden collapse of the suspended layer is
explained by considering a force balance between gravity and solid-like
friction with walls. The present results would be relevant to applications for
mixing a fluid with grains.

\section*{Methods}

\subsection*{Experimental}

The experiment is performed with a cell composed of acrylic plates of
thickness 3 mm. The height $2H$ is fixed to 162 mm, whereas the thickness $L$
and the slope $\alpha$ (that defines the angle $\theta$) range from 3 to 56
mm, and from 60 to 85 deg., respectively. The outlet width $2w$ of the hopper
is varied from 1.5 to 3 mm. The average diameter $d$ of beads is changed from
30 to 100 $\mu$m (GLB-30, GLB-40, GLB-60, GLB-100, Assoc. Powder Process Ind.
and Eng., Japan). We observe the transport of beads through the outlet by
taking movies using a video camera (Canon, iVHS HF S21) either from the front
or lateral side of the cell.

\subsection*{Theory}

The force $F_{\text{fric}}$ can be estimated by noting that the pressure
inside the granular medium is in the hydrostatic regime (the sand layer is not
thick enough for Janssen's model to be valid \cite{Jansen,Duran1997}): the
pressure inside the granular medium at position $z$ measured from the outlet
is simply given by $p=p_{0}+\rho g(h_{t}^{c}-z)\simeq\rho g(h_{t}^{c}-z)$. The
friction force is acting from the two vertical walls and the other two walls
with slope $\alpha$. We consider a simple case $L\gg H$, in which the latter
contribution becomes dominant. For a wall element $Ldl$ ($dl\sin\theta=dz$) at
the the height $z$, the normal force is given by $\rho g(h_{t}^{c}-z)$: this
element is subject to the friction force $\mu\rho g(h_{t}^{c}-z)Ldl$ with
$\mu$ the coefficient of the maximal static friction along the slope
$\alpha=\tan\theta$ and thus the vertical component is given by $\mu\rho
g(h_{t}^{c}-z)Ldz$. For the layer located between the position $h_{b}^{c}$ to
$h_{t}^{c}$ the total friction $F_{\text{fric}}$ applied by the two walls of
slope $\alpha$ is given by $F_{\text{fric}}$ $=2\mu\rho gL\int_{h_{b}^{c}%
}^{h_{t}^{c}}(h_{t}^{c}-z)dz$, which gives the expression used in deriving Eq.
(\ref{eq3}).


\section*{Acknowledgements}

This research was partly supported by Grant-in-Aid for Scientific Research
(A)\ (No. 24244066) of JSPS, Japan.

\section*{Author contributions statement}

K.O and Y.Y. conceived the experiments, Y.Y. conducted most of the
experiments, H.Z.T. also conducted experiments on the phase diagram, K.O and
Y.Y. analyzed the results, Y.Y. and K.O. prepared the figures and graphs, K.O
wrote the manuscript. All authors reviewed the manuscript.

\section*{Additional information}

Competing financial interests: The authors declare no competing financial interests.


\end{document}